\documentclass[pss]{wiley2sp} 
\usepackage{amsmath,amsfonts,amssymb}

\usepackage{bm}              
\usepackage{w-greek}         
\usepackage{units}

\usepackage[bookmarks=true,letterpaper=true,colorlinks=true,urlcolor=blue,linkcolor=blue,citecolor=blue]{hyperref} 
\def\copyrightmark{{\copyrightmarkfont}}%
\journalname{submitted to physica status solidi}

\begin{document}

\title{Theoretical spectroscopy techniques applied to graphene EELS and optics
}

\titlerunning{Theoretical spectroscopy techniques applied to graphene EELS and optics}

\author{%
  D. J. Mowbray\textsuperscript{\Ast,\textsf{\bfseries 1,2}},
  P. Ayala\textsuperscript{\textsf{\bfseries 3}},
  V. Despoja\textsuperscript{\textsf{\bfseries 2}},
  T. Pichler\textsuperscript{\textsf{\bfseries 3}}, and
  A. Rubio\textsuperscript{\textsf{\bfseries 1,2,4}}
}

\authorrunning{D. J. Mowbray et al.}

\mail{e-mail
  \textsf{duncan.mowbray@gmail.com}, Phone:
  +34 943 01 8288, Fax: +34 943 01 8390}

\institute{%
  \textsuperscript{1}\,Nano-bio Spectroscopy Group and ETSF Scientific Development Centre, Depto.~F{\'{\i}}sica de Materiales, Universidad del Pa{\'{\i}}s Vasco, Av.~Tolosa 72, E-20018 San Sebasti{\'{a}}n, Spain\\
  \textsuperscript{2}\,Centro de F{\'{\i}}sica de Materiales CSIC-UPV/EHU-MPC and Donostia Internal Physics Center DIPC, Paseo Manuel de Lardizabal, 4, E-20018 San Sebasti{\'{a}}n, Spain\\
  \textsuperscript{3}\,Faculty of Physics, University of Vienna, Strudlhofgasse 4, A-1090 Vienna, Austria\\
  \textsuperscript{4}\,Fritz-Haber-Institut der Max-Planck-Gesellschaft, Berlin, Germany}

\received{30 April 2011}


\abstract{%
%
%
%
A thorough understanding of the electronic structure is a necessary first step for the design of nanoelectronics, chemical/bio-sensors, electrocatalysts, and nanoplasmonics using graphene. As such, theoretical spectroscopic techniques to describe both direct optical excitations and collective excitations of graphene are of fundamental importance. Starting from density functional theory (DFT) we use the time dependent linear response within the random phase approximation (TDDFT-RPA)  to describe the loss function $-\Im\{\varepsilon^{-1}(\textbf{q},\omega)\}$ for graphene.  To ensure any spurious interactions between layers are neglected, we employ both a radial cutoff of the Coulomb kernel, and extra vacuum directly at the TDDFT-RPA level.
}

%
%

\maketitle   

\section{Introduction}

\begin{figure}[tbp]
\centering
\sidecaption
 \includegraphics*[width=0.35\linewidth]{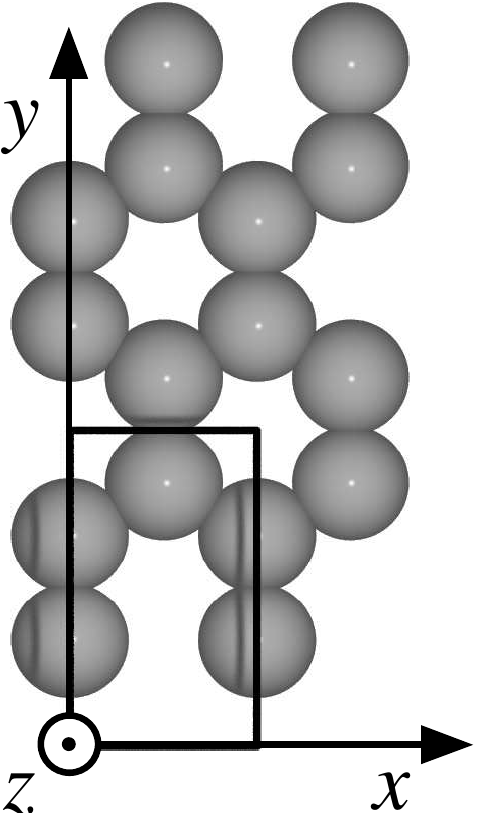}
\caption{%
Schematic of the orthorhombic graphene unit cell repeated twice in the surface plane.  The $x$-direction corresponds to the zigzag direction or circumference of a zigzag SWNT, while the $y$-direction corresponds to the armchair direction or circumference of an armchair SWNT.  The $z$-direction is normal to the graphene surface.
}\label{Graphene_Schematic}
\end{figure}

Understanding of the momentum dependent response function is one of the major challenges in solid state spectroscopy.  One typical example is the analysis of the two particle excitation spectra as determined by the loss function of inelastic electron scattering. In contrast to direct optical excitations, such electron energy loss spectroscopy gives a direct probe of both collective longitudinal excitations (inter and intra-band plasmons), as well as single particle excitations. The momentum dependence of these plasmon excitations as well as single particle excitation gives a measure of the combined dispersion of the valence and conduction bands. Additionally, one can extract information on the screening and dispersion of free charge carriers in low dimensional systems.

To address these questions, we have performed density functional theory (DFT) calculations within the projector augmented wavefunction (PAW) method, and employed the Casida method to calculate the non-interacting density-density response function within the random-phase approximation (TDDFT-RPA), to obtain the loss function $-\Im\{\varepsilon^{-1}(\textbf{q},\omega)$ \cite{Angel,Oni2002RMP}.  Using this methodology, we have calculated the loss function for graphene along the armchair or $y$-direction, as shown schematically in Fig.~\ref{Graphene_Schematic}.  

\section{Methodology}\label{Methodology}

All DFT calculations were performed using the real-space
projector augmented wavefunction (PAW) method code \textsc{gpaw} \cite{GPAW}, with
a grid spacing of 0.2~\AA, and the PBE exchange correlation (xc)-functional \cite{PBE}.  An electronic temperature of $k_B T \approx$ 0.05~eV was used to obtain the occupation of the Kohn-Sham orbitals, with all energies extrapolated to $T = 0$ K, and one unoccupied band per C atom included to improve convergence.

Structural minimization was performed within the Atomic Simulation Environment (ASE)\cite{ASE}, until a maximum force below 0.05~eV/\AA\ was obtained.  An orthorhombic 2.46~\AA\ $\times$ 4.26~\AA\ $\times$ 8.00~\AA\ supercell was employed, consisting of four C atoms, as depicted in Fig.~\ref{Graphene_Schematic}. Non-periodic boundary conditions were enforced in the $z$-direction normal to the graphene surface, so that both the electron density and Kohn-Sham wavefunctions $\rightarrow $ 0 as $z \rightarrow$ 0 or $z\rightarrow L$, where $L$ is the length of the unit cell in the $z$-direction. A Monkhorst-Pack $k$-point sampling of 25 $k$-points along the zigzag direction, and 15 $k$-points along the armchair direction of the graphene surface was employed, yielding a longitudinal momentum transfer resolution $\Delta q$ of 0.102~\AA$^{-1}$ and 0.098~\AA$^{-1}$ respectively.  14 unoccupied bands per C atom were converged in the self-consistent calculation, which was ultimately found to be more than sufficient to converge the loss function for energies up to 50~eV, as shown in Fig.~\ref{Fig1}.

Calculations of the loss function have been performed within the Casida methodology \cite{Angel}, employing time dependent density functional theory within the random phase approximation (TDDFT-RPA), as recently implemented within \textsc{gpaw} \cite{TDDFT}.  Within this framework the non-interacting density-density function $\chi_{\textbf{GG}'}^0(\textbf{q},\omega)$ for momentum transfer $\textbf{q}$ at energy $\omega$ is given by
\begin{eqnarray}
\chi_{\textbf{G}\textbf{G}'}^0(\textbf{q},\omega) &=& \frac{1}{\Omega}\sum_{\textbf{k}}\sum_{n,n'}\frac{f_{n\textbf{k}} - f_{n'\textbf{k}+\textbf{q}}}{\omega + \varepsilon_{n\textbf{k}} - \varepsilon_{n'\textbf{k}+\textbf{q}} + i\gamma}\nonumber\\
&&\times\int_{\Omega_{\textrm{Cell}}}\!\!\!d\textbf{r}\psi^*_{n\textbf{k}}(\textbf{r})e^{-i(\textbf{q}+\textbf{G})\cdot\textbf{r}}\psi_{n'\textbf{k}+\textbf{q}}(\textbf{r})\nonumber\\
&&\times\int_{\Omega_{\textrm{Cell}}}\!\!\!d\textbf{r}'\psi^*_{n\textbf{k}}(\textbf{r}')e^{i(\textbf{q}+\textbf{G}')\cdot\textbf{r}'}\psi_{n'\textbf{k}+\textbf{q}}(\textbf{r}').\label{chi0}
\end{eqnarray}
Here the sum is over reciprocal lattice vectors $\textbf{k}$ and band numbers $n$ and $n'$, with $\varepsilon_{n\textbf{k}}$ the eigenenergy of the $n^{\textrm{th}}$ band at $\textbf{k}$, $f_{n\textbf{k}}$ the Fermi-Dirac occupation of the $n^{\textrm{th}}$ band at $\textbf{k}$, $\gamma$ the peak broadening, $\Omega$ the volume of the supercell, $\textbf{G}$ and $\textbf{G}'$ the reciprocal unit cell vectors, and $\psi_{n\textbf{k}}(\textbf{r})$ the real-space Kohn-Sham wavefunctions for the $n^{\textrm{th}}$ band with reciprocal lattice-vector $\textbf{k}$.


Including local field effects, we may write the inverse macroscopic dielectric function \(\varepsilon^{-1}(\textbf{q},\omega)\) within the random phase approximation (RPA) in terms of the non-interacting density-density response function $\chi_{\textbf{GG}'}^0(\textbf{q},\omega)$ as
\begin{eqnarray}
\varepsilon^{-1}(\textbf{q},\omega) &\approx& \left.\left[\delta_{\textbf{G}\textbf{G}'} - v_{\textbf{G}}(\textbf{q})\chi_{\textbf{GG}'}^0(\textbf{q},\omega)\right]^{-1}\right|_{\textbf{G} = \textbf{G}' = 0},\label{3}
\end{eqnarray}
where \(\delta_{\textbf{GG}'}\) is the Kronecker delta, and $v_{\textbf{G}}(\textbf{q})$ is the Fourier transform of the Coulomb kernel.  It should be noted that the inclusion of exchange and correlation effects in $v_{\textbf{G}}(\textbf{q})$ at the LDA level adds a minor correction to the present results, as already shown for the case of graphite \cite{Pichler,Kramberger08PRL}.

As discussed in Ref.~\cite{RadialCutoff}, for a 3D periodic system with translational invariance, the Coulomb kernel is
\begin{eqnarray}
v^{3\textrm{D}}_{\textbf{G}}(\textbf{q}) &=& \iiint d\textbf{r}\frac{e^{i(\textbf{q}+\textbf{G})\cdot\textbf{r}}}{\|\textbf{r}\|} = \frac{4\pi}{\|\textbf{q}+\textbf{G}\|^2}.
\end{eqnarray}

However, for a system which is periodic in only two dimensions, such as a bulk slab or graphene, interactions between periodic images in a TDDFT-RPA calculation may be significant due to the long-range behaviour of $v^{3\textrm{D}}$.  This will be the case even for systems with sufficient vacuum to converge the electron density at the DFT level.  On the other hand, image---image interactions are included at the TDDFT-RPA level only through $v^{3\textrm{D}}$.  This motivates us to introduce a 2D periodic Coulomb kernel, $v^{2\textrm{D}}$, which is both translationally invariant and zero for $|z| > R$, where $R$ is the ``radial cutoff'' for the Coulomb kernel.  In this case
\begin{eqnarray}
v^{2\textrm{D}}_{\textbf{G}}(\textbf{q}) &=& \int_{-R}^R dz\iint dx dy \frac{e^{i (\textbf{q}+\textbf{G})\cdot(\textbf{x}+\textbf{y}+\textbf{z})}}{\sqrt{x^2+y^2+z^2}}\nonumber\\
 &=& \frac{4\pi}{\|\textbf{q}+\textbf{G}_\|\|} \int_0^R \cos(G_z z) e^{-\|\textbf{q}+\textbf{G}_\|\|z} dz\nonumber\\
&=& \frac{4\pi\left[1 + e^{-\|\textbf{q}+\textbf{G}_\|\|R}\left[\frac{G_z \sin G_z R}{\|\textbf{q}+\textbf{G}_\|\|} - \cos G_zR\right]\right]}{\|\textbf{q}+\textbf{G}\|^2}\nonumber.
\end{eqnarray}
Employing the suggested choice of $R = L/2$ \cite{RadialCutoff}, since $G_z = n \frac{2\pi}{L}$, where $n \in \mathbb{Z}$, we find
\begin{eqnarray}
v^{2\textrm{D}}_{\textbf{G}}(\textbf{q}) &=& \frac{4\pi}{\|\textbf{q}+\textbf{G}\|^2}\left[1 - e^{-\|\textbf{q}+\textbf{G}_\|\|L/2}\cos n\pi\right].\label{v2D}
\end{eqnarray}
From Eqn.~\eqref{v2D} we clearly see that for $L \gg 2/q$ or $q \gtrsim$ 1~\AA$^{-1}$, $v^{2\textrm{D}} \rightarrow v^{3\textrm{D}}$.  

Alternatively, we may introduce further regions of vacuum separating the images directly at the TDDFT-RPA level.  Since in the added vacuum regions the matrix elements for the occupied Kohn-Sham wavefunctions are zero, the inclusion of extra vacuum in Eqn.~\eqref{chi0} only enters into the non-interacting density-density response function through the unit cell volume $\Omega$ and the reciprocal unit cell vectors $\textbf{G}$.  We may thus introduce extra unit cells of vacuum, or ``zero padding'' in the non-periodic direction, by doubling or tripling $L$ when computing the set of $\textbf{G}$ vectors to include at the TDDFT-RPA level.  
In this way, increasing the length of the unit cell in the non-periodic direction through the inclusion of vacuum effectively increases the density of sampling of the reciprocal unit cell.  

Finally, the quantities of fundamental interest are the  loss function \(-\Im\{\varepsilon^{-1}(\textbf{q},\omega)\}\), the adsorption or imaginary part of the dielectric function \(\Im\{\varepsilon(\textbf{q},\omega)\}\), and the real part of the dielectric function \(\Re\{\varepsilon(\textbf{q},\omega)\}\), which may be obtained from Eqn.~\eqref{3}.  




\section{Results \& Discussion}\label{Results}

\begin{figure}[!t]
\centering
\sidecaption
\includegraphics*[height=4.5in]{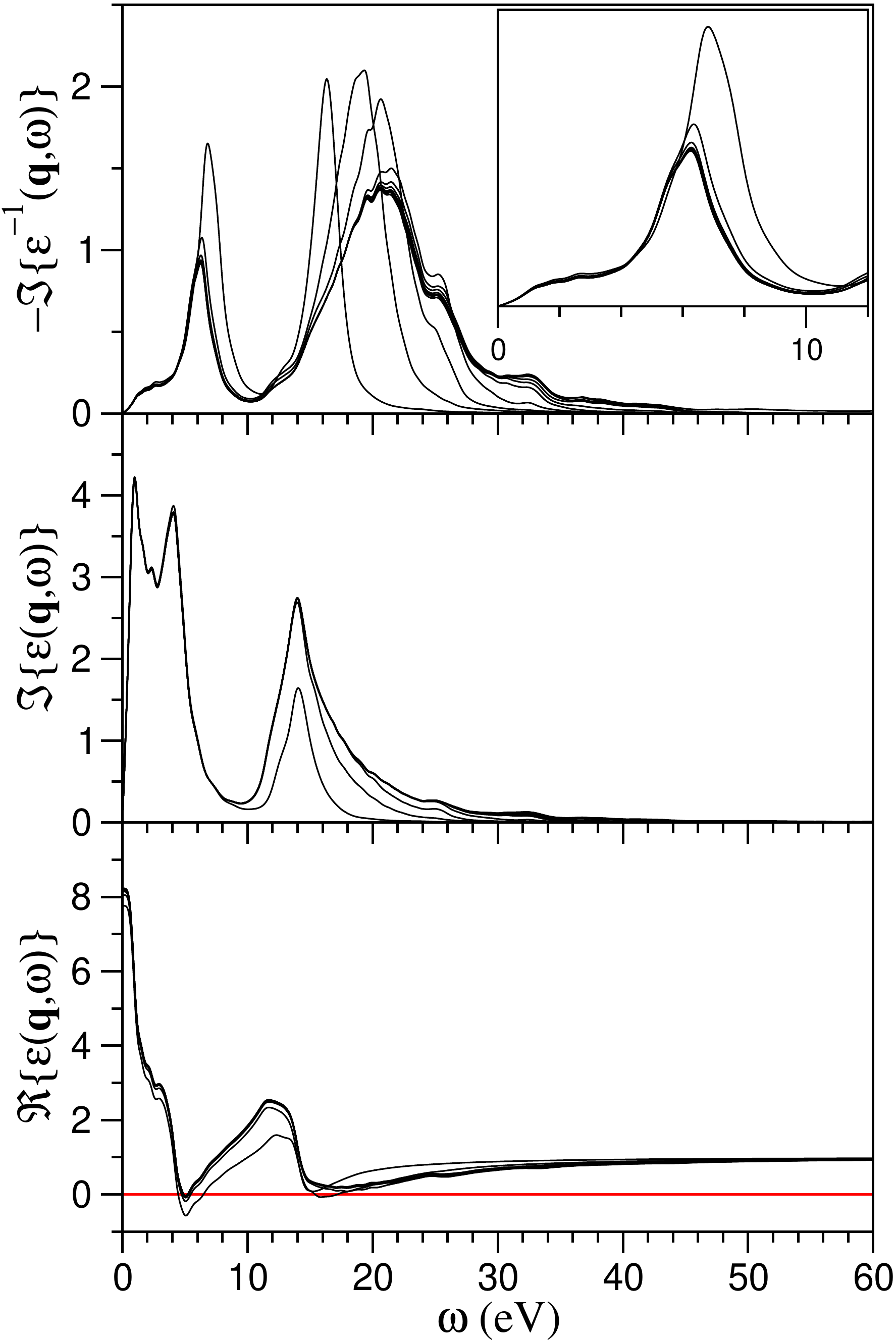}
\caption{Convergence with respect to the number of unoccupied bands $n_\textrm{unocc}$ of the calculated loss function $-\Im\{\varepsilon^{-1}(\textbf{q},\omega)\}$, imaginary part of the dielectric function $\Im\{\varepsilon(\textbf{q},\omega)\}$, and real part of the dielectric function $\Re\{\varepsilon(\textbf{q},\omega)\}$  versus energy $\omega$ in~eV for momentum transfer $\|\textbf{q}\| \approx$ 0.1~\AA$^{-1}$ along the armchair direction of graphene.  Contributions from including  $n_{\textrm{unocc}}$ = 1, 2, 3, 4, 5, 6, 7, 8, 10, and 14 unoccupied bands per C atom are shown. A plane-wave cutoff energy $\varepsilon_{\textrm{cut}}$ = 60~eV, corresponding to 89 $\textbf{G}$ vectors was used.}\label{Fig1}
\end{figure}

\begin{figure}[!t]
\centering
\sidecaption
\includegraphics*[height=4.5in]{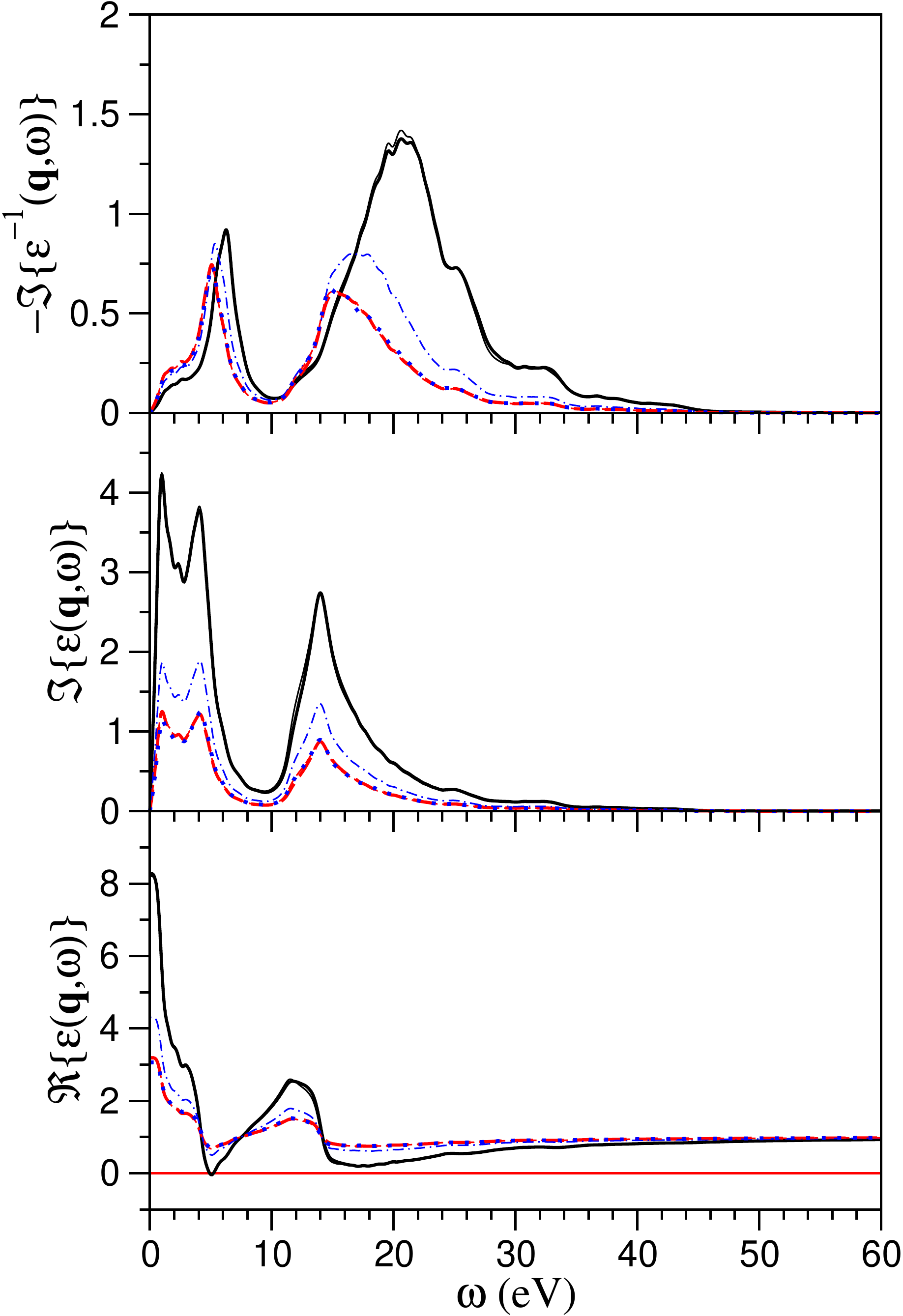}
\caption{Influence of neighbouring image interactions on the calculated loss function $-\Im\{\varepsilon^{-1}(\textbf{q},\omega)\}$, imaginary part of the dielectric function $\Im\{\varepsilon(\textbf{q},\omega)\}$, and real part of the dielectric function $\Re\{\varepsilon(\textbf{q},\omega)\}$  versus energy $\omega$ in~eV for momentum transfer $\|\textbf{q}\| \approx$ 0.1~\AA$^{-1}$ along the armchair direction of graphene, and including $n_{\textrm{unocc}}$ = 8 bands/atom.  Results are shown for $\varepsilon_{\textrm{cut}}$ = 15~eV with 8~\AA\ of vacuum (------), 16~\AA\ of vacuum ({\color{blue}{-- $\cdot$ --}}), and 24~\AA\ of vacuum ({\color{blue}{$\cdots \cdots$}}), corresponding to 11, 25, and 37 $\textbf{G}$ vectors respectively, and for $\varepsilon_{\textrm{cut}}$ = 60~eV with 8~\AA\ of vacuum, corresponding to 89 $\textbf{G}$ vectors, both with ({\textbf{\color{red}{-- -- --}}}) and without ({\textbf{------}}) applying a radial cutoff $R$ = 4~\AA\ to the Coulomb kernel.}\label{Fig2D}
\end{figure}

\begin{table}
\centering
\caption{Percentage error for the $f$-sum rule $\Delta f$, and convergence with the number of unoccupied bands $n_{\textrm{unocc}}$ per C atom.}\label{f-sum-rule}
%
\begin{tabular}[tbhp]{cccc}\hline
$n_{\textrm{unocc}}$ & $\Delta f$ & $n_{\textrm{unocc}}$ & $\Delta f$ \\
(bands/atom) & (\%) & (bands/atom) & (\%) \\\hline
1 & 66\%&6 &18\%\\
2 &40\%&7 &16\%\\
3 &28\%&8 &15\%\\
4 &23\%&10 &14\%\\
5 &20\%&14 &12\%\\\hline
\end{tabular}
\end{table}

To test the convergence of the TDDFT-RPA calculated dielectric function with respect to the number of unoccupied bands included within the calculation $n_{\textrm{unocc}}$, we make use of the $f$-sum rule,
\begin{eqnarray}
\int_0^\infty d\omega \omega \Im\{\varepsilon(\omega)\} &=& \frac{2\pi^2 N_e}{\Omega},
\end{eqnarray}
where $N_e$ is the number of electrons in the unit cell of volume $\Omega$.  The percentage error in the adherence of the imaginary part of the calculated dielectric function to the sum rule, $\Delta f$, for $n_{\textrm{unocc}} =$ 1 to 14 unoccupied bands per C atom is shown in Table~\ref{f-sum-rule}.  For $n_{\textrm{unocc}} =$ 8 bands/atom, the $f$-sum rule is quite well satisfied, considering that  
the integral has been truncated at $\omega_{\textrm{max}} = $ 60 eV, and the approximation $d\omega \approx \Delta\omega = $ 0.02 eV.  It should also be noted that the $f$-sum rule is not satisfied exactly when using non-local pseudopotentials, although the correction is within the errors given in Table~\ref{f-sum-rule} \cite{Oni2002RMP}.

\begin{figure}[tbp]
\centering
\sidecaption
\includegraphics*[width=\linewidth]{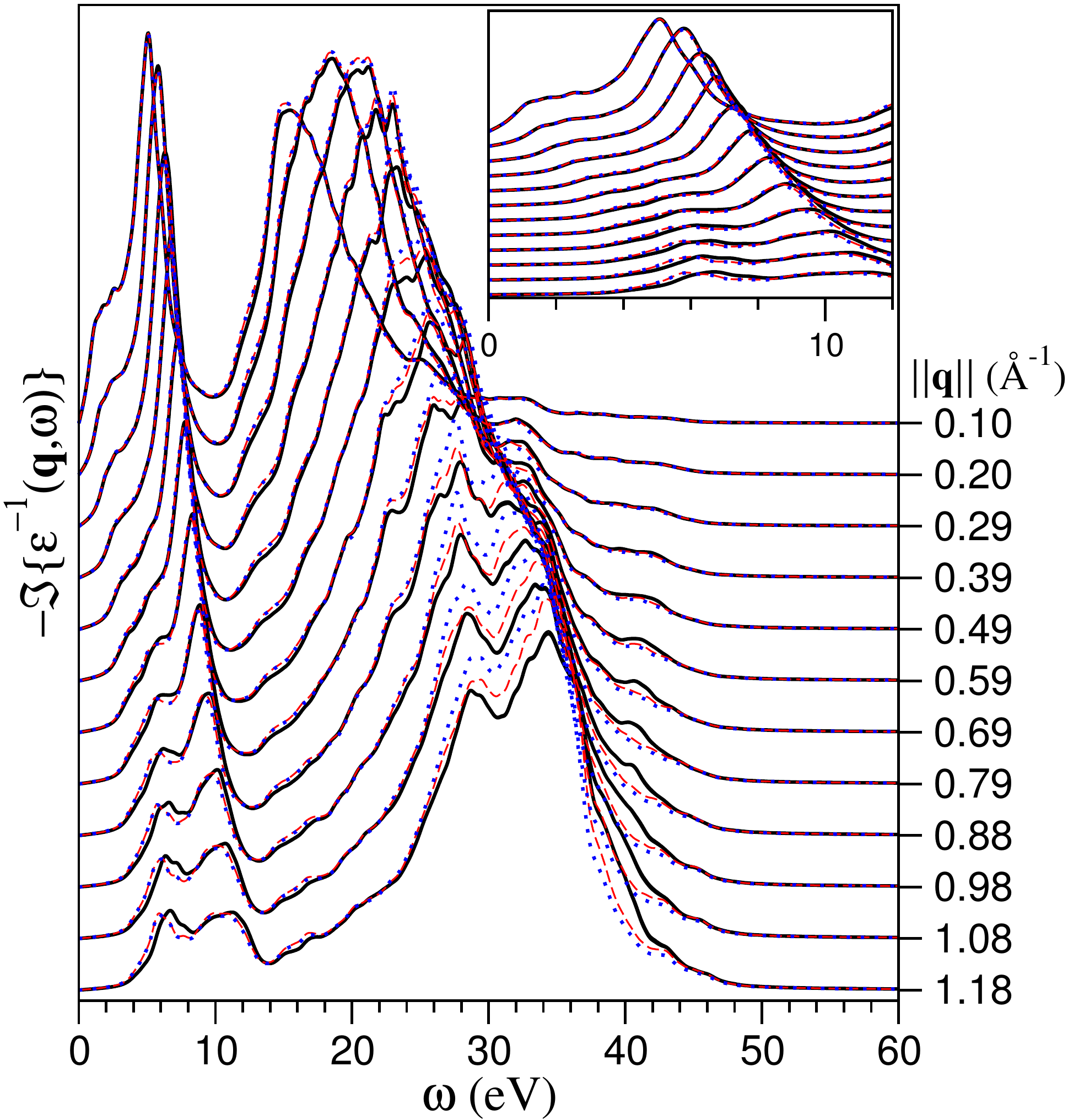}
\caption{Convergence with respect to the plane-wave cutoff energy $\varepsilon_\textrm{cut}$ for the calculated loss function $-\Im\{\varepsilon^{-1}(\textbf{q},\omega)\}$  versus energy $\omega$ in~eV and momentum transfer $\|\textbf{q}\|$ in~\AA$^{-1}$ along the armchair direction of graphene, including $n_{\textrm{unocc}}$ = 8 bands/atom, and a radial cutoff $R =$ 4 \AA.  Results for $\varepsilon_\textrm{cut}$ = 15 ({\color{blue}{$\cdots \cdots$}}), 30 ({\color{red}{-- -- --}}), 60~eV (------), and 90~eV ({\textbf{------}}), corresponding to 11, 27, 89, and 149 $\textbf{G}$ vectors respectively, are shown.}\label{Fig5}
\end{figure}

\begin{figure}[tbp]
\centering
\sidecaption
\includegraphics*[width=\linewidth]{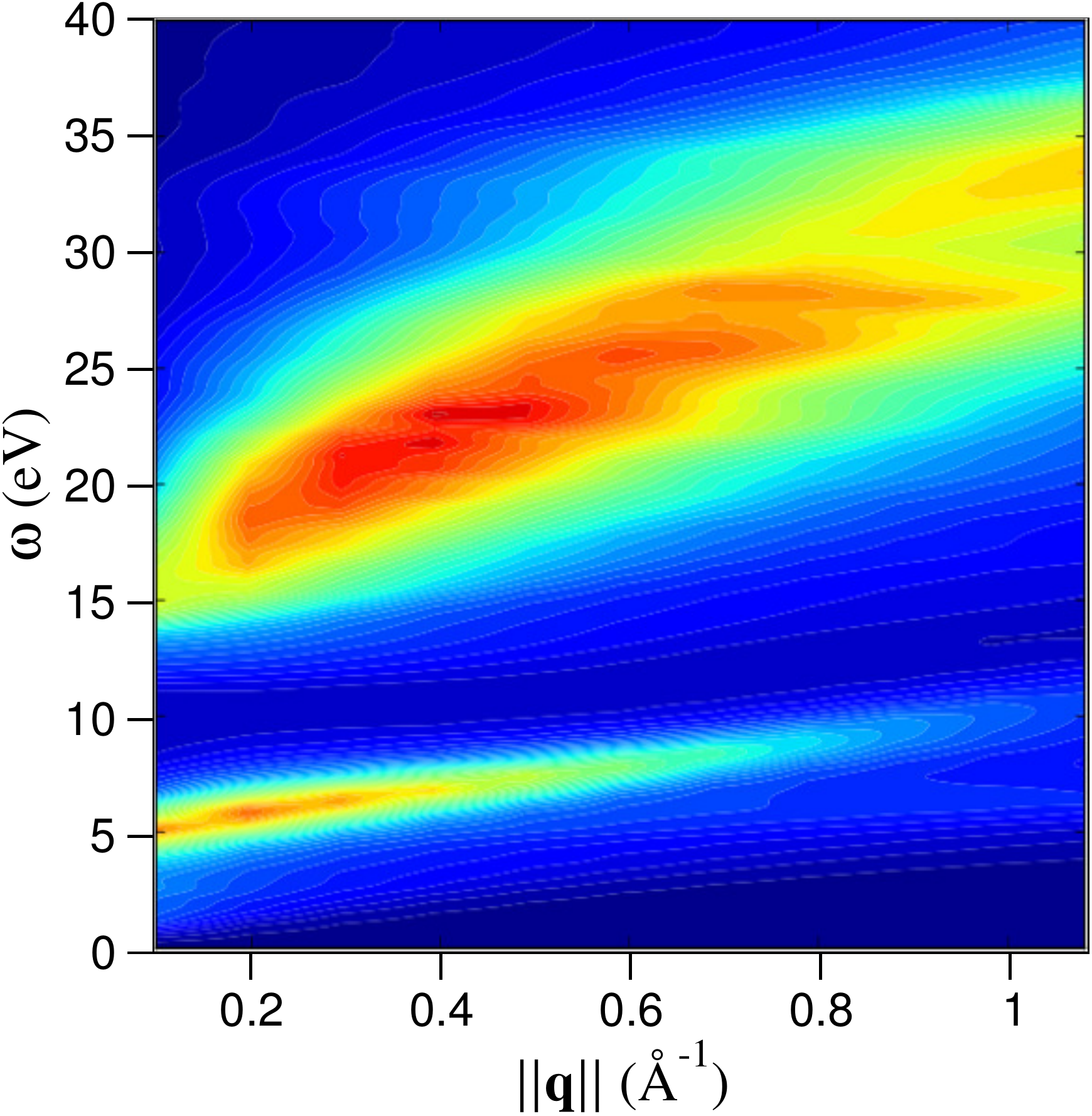}
\caption{Intensity of the calculated loss function $-\Im\{\varepsilon^{-1}(\textbf{q},\omega)\}$ as a function of energy $\omega$ in~eV and momentum transfer $\|\textbf{q}\|$ in~\AA$^{-1}$ along the armchair direction of graphene, with $n_{\textrm{unocc}}$ = 8 bands/atom, $\varepsilon_{\textrm{cut}}$ = 60~eV, corresponding to 89 $\textbf{G}$ vectors, and using a radial cutoff $R$ = 4~\AA.  }\label{Fig3}
\end{figure}

  
As a more direct test, we plot in Fig.~\ref{Fig1} the loss function and both the real and imaginary parts of the dielectric function for $n_{\textrm{unocc}}$ between 1 and 14 bands per C atom.  We consider the limit of low momentum ($\|\textbf{q}\| \approx$ 0.1 \AA$^{-1}$ in the armchair direction) since in this limit $\chi_{\textbf{G}\textbf{G}'}^0$ will be rather insensitive to number of $\textbf{G}$ vectors included. 

Figure \ref{Fig1} clearly shows that (1) the dielectric function converges faster than the loss function with respect to $n_{\textrm{unocc}}$, (2) for $n_{\textrm{unocc}} =$ 8 bands/atom we have convergence up to about 50 eV, and (3) with only 2--3 unoccupied bands per C atom the loss function is converged semi-quantitatively up to 10 eV, as shown in the inset.  Combined with the values for the $f$-sum rule error in Table~\ref{f-sum-rule}, this suggests that $n_{\textrm{unocc}} = $ 8 bands/atom is sufficient to describe the dielectric response of graphene-like materials up to 50 eV.  Further, a near quantitative description of the behaviour of the low energy $\pi$ plasmons is obtained by including only 2--3 unoccupied bands per C atom.

However, although both the electron density and Kohn-Sham wavefunction have been set to zero at $z = 0$ and $z=L$ through the use of non-periodic boundary conditions at the DFT level, interactions between periodic images in the $z$-direction are included through the Coulomb kernel in Eqn.~\eqref{3}.  Figure \ref{Fig2D} shows how much neighbouring image interactions contribute to the loss and dielectric functions for low momentum transfers.  By either applying a radial cutoff of $R=$ 4 \AA\ using $v^{2\textrm{D}}$ or increasing the unit cell length $L$ to 24 \AA\ at the TDDFT-RPA level, we obtain the same converged loss and dielectric functions, with image---image interactions removed.  However, as the addition of extra unit cells of vacuum, even at only the TDDFT-RPA level, increases the size of the \(\chi_{\textbf{G}\textbf{G}'}^0\) matrix, we shall employ a radial cutoff of $R = $ 4 \AA\ from hereon.  

Figure \ref{Fig5} shows the convergence of the loss function dispersion with respect to the plane-wave energy cutoff $\varepsilon_{\textrm{cut}}$, and the corresponding number of $\textbf{G}$ vectors included in the calculation.  As mentioned above, for low to medium momentum transfers (\(\|\textbf{q}\| \lesssim\) 0.8 \AA$^{-1}$) only a few $\textbf{G}$ vectors (10---30) are required to converge the loss function.  Further, even at high momentum transfers (\(\|\textbf{q}\| \gtrsim\) 1 \AA$^{-1}$), the differences between \(\varepsilon_{\textrm{cut}} =\) 15 and 90 eV lie mainly in the intensities rather than the peak positions.  This is especially true if we restrict consideration to the perhaps most interesting low energy $\pi$ plasmon regime.

The intensity of the converged loss function for graphene is plotted in Fig.~\ref{Fig3} for $n_{\textrm{unocc}} = $ 8 bands/atom, 89 \textbf{G} vectors, and a radial cutoff $R = $ 4 \AA.  We clearly see the near linearly dispersive $\pi$ plasmon at 5---10 eV, the non-dispersive $\pi$ plasmon at about 5.5 eV, and the dispersing $\sigma+\pi$ plasmon between 15--30 eV, from experiment \cite{Kramberger08PRL}.   We also find that the optical limit $\|\textbf{q}\|\rightarrow 0^+$ for graphene is well reproduced \cite{PhysRevLett.106.046401}.

\section{Conclusions}\label{Conclusions}

In conclusion, we have extended the TDDFT-RPA implementation within \textsc{gpaw} to employ both a radial cutoff of the Coulomb kernel $v^{2\textrm{D}}$ for 2D periodic systems, and include extra unit cells of vacuum at the TDDFT-RPA level. We find these spurious image---image interactions have a significant impact on the calculated loss function for isolated systems at low momentum transfer (\(\|\textbf{q}\| \lesssim\) 0.5 \AA$^{-1}$), as demonstrated for graphene.  We also find for carbon systems, including about 3 unoccupied bands per atom is sufficient to obtain a near quantitative description of the loss function up to 10 eV.  

These results are particularly important in the area of nanoplasmonics, for the description of the low energy free-charge carrier plasmons induced by electrostatic or potassium doping.  Such systems can be well described within a simple RPA model by shifting the Fermi level rigidly, as we will show in future work.

\begin{acknowledgement}
D.J.M. acknowledges funding through the Spanish ``Juan de la Cierva'' program (JCI-2010-08156).  We acknowledge funding by Spanish MEC (FIS2010-21282-C02-01), ``Grupos Consolidados UPV/EHU del Gobierno Vasco'' (IT-319-07), ACI-Promociona (ACI2009-1036), and the Ikerbasque Foundation.   The European Theoretical Spectroscopy Facility is funded through ETSF-I3 (Contract Number 211956).
  \end{acknowledgement}

%
\bibliographystyle{arXiv}

\end{document}